\documentclass[12pt]{elsart}
\usepackage{graphicx,amssymb}
\journal{Physica B}
\newcommand{\sect}[1]{\setcounter{equation}{0}\section{#1}}
\newcommand{\bfm}[1]{\mbox{\boldmath${#1}$}}

\begin{document}
\begin{frontmatter}
\title{Travelling waves in a mixture of gases
with bimolecular reversible reactions}
\author{A. Rossani}\footnote{alberto.rossani@polito.it; corresponding author.}$^{\hspace{-1mm},\, \rm
a}$
\author{A.M. Scarfone}\footnote{antonio.scarfone@polito.it}$^{\hspace{-1mm},\, \rm a,b}$
\address{$^{\rm a}$Istituto
Nazionale di Fisica della Materia and Dipartimento di Fisica\\
Politecnico di Torino, Corso Duca degli Abruzzi 24, 10129 Torino,
Italy} \address{$^{\rm b}$Istituto Nazionale di Fisica Nucleare,
Universit{\'a} di Cagliari,\\ Cittadella Universitaria, 09042
Monserrato Cagliari, Italy}
\date{\today}
\begin {abstract}
Starting from the kinetic approach for a mixture of reacting
gases whose particles interact through elastic scattering and a
bimolecular reversible chemical reaction, the equations that
govern the dynamics of the system are obtained by means of the
relevant Boltzmann-like equation. Conservation laws are
considered. Fluid dynamic approximations are used at the Euler
level to obtain a close set of PDEs for six unknown macroscopic
fields. The dispersion relation of the mixture of reacting gases
is explicitly derived in the homogeneous equilibrium state. A set
of ODE that governs the propagation of a plane travelling wave is
obtained using the Galilei invariance. After numerical
integration some solutions, including the well-known Maxwellian
and the hard spheres cases, are found for various meaningful
interaction laws. The main macroscopic observables for the gas
mixture such as the drift velocity, temperature, total
density, pressure and its chemical composition are shown.\\
\end {abstract}
\end{frontmatter}
\sect{Introduction} The study of chemically reacting gases has
been dealt with using the kinetic theory since the 1940s
\cite{Prigogine}, and a large amount of literature has been
developed about chemically reacting rarefied flows. This research
field led to applications in the study of combustion \cite{KUO},
detonation (see \cite{CMSZ} and references therein) and
atmospheric chemistry \cite{JAC}, fields in which a precise
knowledge of the dynamics of the reacting gases mixture is required.\\
Due to the increasing complexity that arises from a detailed
microscopic description, there are not many results in literature
and they mainly concern numerical simulations and applications.\\
Starting from the kinetic approach, fluid dynamic approximations
have been constructed as simplified tools for a description, at a
macroscopic level, of the mixture we are dealing with
\cite{XD,BGS}. In order to recall some of the main contributions
in this field, we can quote Ref.s \cite{RM,SK} where the
Chapman-Enskong approximation method was used in the homogeneous
space case for an accurate evaluation of the chemical reaction
rate and, more recently, the Grad method which was applied in both
the homogeneous and non-homogeneous space cases, in order to
calculate the reaction rate and transport coefficients \cite{XD,M}.\\
In a recent paper \cite{RS}, Rossani and Spiga obtained a more
explicit derivation of kinetic equations with a chemical reaction,
and succeeded in completely characterizing the equilibrium
distributions and in proving an H-theorem for them. A
generalization to many level molecules was performed in \cite{GS}.

In the present work, the authors deal with a gas that undergoes to
the following reversible bimolecular reaction
\begin{eqnarray}
A+B\rightleftharpoons C+D \ .\label{reaction}
\end{eqnarray}
It should be mentioned that the kinetic theory governing reaction
(\ref{reaction}) has been the subject of investigation in recent
years. For instance, in \cite{GRS} the kinetic model of a
diatomic gas with both dissociation and recombination reactions
through a transition state was proposed, whereas in \cite{GR} the
extended Boltzmann equation governing reaction (\ref{reaction})
was investigated in the Lorentzian gas limit.\\ In the first part
of this paper, starting from the nonlinear system of kinetic
equations for a mixture of particles $A$, $B$, $C$, and $D$, we
derive the balance equations for the different chemical species as
well as the conservation equations for the macroscopic quantities
such as mass, momentum and total energy. By adopting the Euler
approximation, in the case of a homogeneous equilibrium state, we
analytically derive the dispersion relation for the mixture of
the reacting gases, where we found a relationship like
$k^2\,P(i\,\omega)+\omega^2\,Q(i\,\omega)=0$, where $P$ and $Q$
are polynomials of the first order in $i\,\omega$, with real
coefficients.\\ In the second part, we deal with the study of the
propagation of a plane travelling wave. It can be remarked that,
although many results can be found in literature about this
important fluid dynamic problem for a dissociating gas (see
\cite{VK,Cercignani} and references therein), the explicit
treatment of a travelling wave with bimolecular reversible
reactions is very recent \cite{CMSZ}. \\In the dynamics of
non-reacting gases, viscosity and heat conduction are responsible
for the structure of a travelling wave \cite{Cercignani}. However,
here we make the assumption that relaxation due to elastic
scattering is much quicker than the one due to chemical
interactions. In this way we can neglect viscosity and heat
conduction, so that only the chemical reaction is responsible for
the structure of the travelling wave. Under the previous
hypothesis on relaxation times, we construct a close set of
nonlinear PDEs that govern the propagation
of a plane steady travelling wave through the reacting gas.\\
By taking advantage of the Galilei invariance of the Euler
equations, we can adopt the reference frame in which the
travelling wave is at rest. The resulting ODE system is reduced
to a single nonlinear ODE which is solved numerically for several
meaningful interaction laws, including the case of Maxwell and
hard sphere molecules. Results on the whole gas (drift velocity,
temperature, total density and pressure) as well as its chemical
composition (molar fractions of the four components) are plotted.
The effects of the different choices for the interaction laws on
the structure of the travelling wave are discussed.

\sect{Kinetic equations} Let us consider a mixture of four gases
$A$, $B$, $C$ and $D$, which can interact, apart from all the
elastic collisions, according to the reversible bimolecular
reaction given in (\ref{reaction}). In the sequel, we denote the
particles $A$, $B$, $C$ and $D$ with the index $i=1,\,2,\,3,\,
4$. The distribution functions $f_i({\bfm x},\ {\bfm v},\ t)$ for
particles $i$, endowed with masses $m_{_i}$ and internal energy
$E_i$ (available in the case of reactive interactions), obey the
following system of nonlinear integro-differential Boltzmann
equations (see \cite{RS} for details):
\begin{eqnarray}
\frac{\partial\,f_{_i}}{\partial\,t}+{\bfm
v}\cdot{\bfm\nabla}\,f_{_i}=J_{_i}[{\bfm f}]+Q_{_i}[{\bfm f}] \
,\label{boltzmann}
\end{eqnarray}
where the vector ${\bfm
f}\equiv(f_{_1},\,f_{_2},\,f_{_3},\,f_{_4})$. In Eq.
(\ref{boltzmann}), $Q_{_i}$ are the well known elastic collision
integrals \cite{Cercignani} while $J_{_i}$ are the chemical
collision terms given by:
\begin{eqnarray}
\nonumber J_{_1}[\bfm f](\bfm
v)=\int\!\!\int\Bigg[\frac{\mu_{_{12}}\,V_{_{34}}^{\,\,2}}{\mu_{_{34}}\,V_{_{12}}}\,
I_{_{34}}^{^{12}}\,f_{_3}(\bfm v^\prime)\,f_{_4}(\bfm w^\prime)-
V_{_{12}}\,I_{_{12}}^{^{34}}\,f_{_1}(\bfm
v)\,f_{_2}(\bfm w)\Bigg]\,d\bfm w\,d\bfm\Omega^\prime \ ,\\
\label{j1}
\end{eqnarray}
where $V_{_{12}}=|{\bfm v}-{\bfm w}|$, $V_{_{34}}=|{\bfm
v}^\prime-{\bfm w}^\prime|$,
$\zeta={\bfm\Omega}\cdot{\bfm\Omega}^\prime$, with
${\bfm\Omega}=({\bfm v}-{\bfm w})/V_{_{12}}$,
${\bfm\Omega}^\prime=({\bfm v}^\prime-{\bfm w}^\prime)/V_{_{34}}$
and
\begin{eqnarray}
\nonumber J_{_3}[\bfm f](\bfm
v)=\int\!\!\int\Bigg[\frac{\mu_{_{34}}\,V_{_{12}}^{\,\,2}}
{\mu_{_{12}}\,V_{_{34}}}\, I_{_{12}}^{^{34}}\,f_{_1}(\bfm
v^\prime)\,f_{_2}(\bfm w^\prime)-V_{_{34}}\,I_{_{34}}^{^{12}}
\,f_{_3}(\bfm v)\,f_{_4}(\bfm w)\Bigg]\,d\bfm
w\,d\,\bfm\Omega^\prime \ ,\\ \label{j3}
\end{eqnarray}
where, now, $V_{_{34}}=|{\bfm v}-{\bfm w}|$, $V_{_{12}}=|{\bfm
v}^\prime-{\bfm w}^\prime|$, ${\bfm\Omega}=({\bfm v}-{\bfm
w})/V_{_{34}}$ and ${\bfm\Omega}^\prime=({\bfm v}^\prime-{\bfm
w}^\prime)/V_{_{12}}$. The explicit expressions of the
post-collision velocities $\bfm v^\prime$ and $\bfm w^\prime$,
(which can be found in Ref. \cite{RS}), are omitted here for the
sake of brevity. The integrals $J_{_2}[\bfm f]$ and $J_{_4}[\bfm
f]$ are obtained from $J_{_1}[\bfm f]$ and $J_{_3}[\bfm f]$ by
introducing the following index exchange: $1\rightleftharpoons
2$, $3\rightleftharpoons 4$. In Eq.s (\ref{j1}) and (\ref{j3}),
$I_{_{34}}^{^{12}}\equiv I_{_{34}}^{^{12}}(V_{_{34}},\,\zeta)$
and $I_{_{12}}^{^{34}}\equiv
I_{_{12}}^{^{34}}(V_{_{12}},\,\zeta)$ are the differential
cross-sections of the forward and backward reaction, respectively,
that are related by means of the following microreversibility
relationship:
$\mu_{_{12}}^{\,\,2}\,V_{_{12}}^{\,\,2}\,I_{_{12}}^{^{34}}(V_{_{12}},\zeta)
=\mu_{_{34}}^{\,\,2}\,V_{_{34}}^{\,\,2}\,I_{_{34}}^{^{12}}(V_{_{34}},\zeta)$,
where $\mu_{_{ij}}=m_{_i}\,m_{_j}/M$ are the reduced masses, and
$M=m_{_1}+m_{_2}=m_{_3}+m_{_4}$ (mass conservation) is the total
mass of the reactants/products. The chemical collision terms
satisfy the following important properties \cite{RS}: $\int
J_{_1}[\bfm f]\,d{\bfm v}=\int J_{_2}[\bfm f]\,d{\bfm v}=-\int
J_{_3}[\bfm f]\,d{\bfm v}=-\int J_{_4}[\bfm f]\,d{\bfm
v}=G_{34}-G_{12} \ ,$ where
\begin{eqnarray}
G_{34}=\int\int\int V_{_{34}}\,I_{_{34}}^{^{12}}\,f_{_3}(\bfm v)\,
f_{_4}(\bfm v^\prime)\,d{\bfm v}\,d{\bfm
v}^\prime\,d{\bfm\Omega}^\prime \ ,\\ G_{12}=\int\int\int
V_{_{12}}\,I_{_{12}}^{^{34}}\,f_{_1}(\bfm v) f_{_2}(\bfm
v^\prime)\,d{\bfm v}\,d{\bfm v}^\prime\,d{\bfm\Omega}^\prime \ .
\end{eqnarray}
In the following we derive macroscopic balance laws starting from
the kinetic equation (\ref{boltzmann}).\\ By integrating Eq.
(\ref{boltzmann}) over $d{\bfm v}$ and using the relation $\int
Q_{_i}[\bfm f]\,d\bfm v=0$ \cite{Cercignani}, we obtain the
balance equations for particles $i$:
\begin{eqnarray}
\frac{\partial\,n_{_i}}{\partial\,t}+{\bfm\nabla}\cdot(n_{_i}\,{\bfm
u_{_i}})=S_{_i} \ ,\label{particle}
\end{eqnarray}
where the density $n_{_i}$, the drift velocity ${\bfm u}_{_i}$ of
the species $i$, and the source term $S_{_i}$ due to the chemical
interactions, are given by: $n_{_i}=\int f_{_i}({\bfm v})\,d{\bfm
v}$, ${\bfm u}_{_i}=(1/n_{_i})\int{\bfm v}\,f_{_i}({\bfm
v})\,d{\bfm v}$ and
$S_{_i}=\lambda_{_i}\,\left(G_{_{34}}-G{_{12}}\right)$ with
$\lambda_{_i}=1$ for $i=1,\,2$ and $\lambda_{_i}=-1$ for
$i=3,\,4$.\\ By multiplying Eq. (\ref{boltzmann}) by $m_{_i}$,
$m_{_i}\,{\bfm v}$ and $E_i+m_{_i}\,{\bfm v}^2/2$, by summing
over $i$ and by integrating over $d{\bfm v}$, we obtain the
conservation equations for the mass, momentum and total energy:
\begin{eqnarray}
&&\frac{\partial\,\rho}{\partial\,t}+{\bfm\nabla}\cdot(\rho\,{\bfm
u})=0 \ ,\label{mass}\\
&&\frac{\partial}{\partial\,t}(\rho\,{\bfm
u})+{\bfm\nabla}\cdot(\rho\,{\bfm u}\otimes{\bfm u}+I\!\!P)={\bfm
0} \ ,\label{momentum}\\
&&\frac{\partial\,e}{\partial\,t}+{\bfm\nabla}\cdot\left[\left(e\,I\!\!I+I\!\!P
\right)\cdot{\bfm u}+{\bfm q}_{_{\rm th}}+{\bfm q}_{_{\rm
int}}\right]=0 \ ,\label{totenergy}
\end{eqnarray}
where $I\!\!I$ is the unit tensor. Equations (\ref{mass}),
(\ref{momentum}), and (\ref{totenergy}) were derived by making use
of the property \cite{RS} $\sum_{i=1}^4\int\psi_{_i}({\bfm
v})\Big(J_{_i}[{\bfm f}]+Q_{_i}[{\bfm f}]\Big)\,d{\bfm v}=0$ which
is valid for $\psi_i=m_i$, $\psi_{_i}=m_{_i}\,{\bfm v}$ (three
components), and $\psi_{_i}=E_{_i}+m_{_i}\,{\bfm v}^2/2$ (mass,
momentum, and total energy conservation). In Eq.s
(\ref{mass})-(\ref{totenergy}) $\rho=\sum_i m_{_i}\,n_{_i}$ is the
total density and ${\bfm u}=(1/\rho)\,\sum_i m_{_i}\,n_{_i}\,{\bfm
u}_{_i}$ is the drift velocity of the mixture. We also posed
$I\!\!P=\sum_i m_{_i}\int({\bfm v}-{\bfm u})\otimes({\bfm
v}-{\bfm u})\,f_{_i}({\bfm v})\,d{\bfm v}$, ${\bfm q}_{_{\rm
th}}={1\over2}\,\sum_i m_{_i}\int({\bfm v}-{\bfm u})^2\,({\bfm
v}-{\bfm u})\,f_{_i}({\bfm v})\,d{\bfm v}$ and ${\bfm q}_{_{\rm
int}}=\sum_i E_{_i}\int({\bfm v}-{\bfm u})\,f_{_i}({\bfm
v})\,d{\bfm v}=\sum_i E_{_i}\,n_{_i}\,({\bfm u}_{_i}-{\bfm u})$
which represent the stress tensor, the thermal energy flux and
the internal energy flux, respectively. Finally,
$e=\frac{1}{2}\,\rho\,{\bfm u}^2+e_{_{\rm th}}+e_{_{\rm int}}$ is
the total energy density where $e_{_{\rm
th}}=\frac{3}{2}\,n\,\chi\,T={1\over2}\,{\rm tr}\,I\!\!P$ and $
e_{_{\rm int}}=\sum_iE_{_i}\,n_{_i}$ are the thermal energy
density ($n=\sum_i n_{_i}$ is the total density, $\chi$ is the
Boltzmann constant and $T$ is the absolute
temperature) and the internal energy density, respectively.\\
Equations (\ref{particle}), (\ref{momentum}) and
(\ref{totenergy}) are exact but do not constitute a closed
system. After suitable assumptions, this system can be made close
in order to obtain a simplified tool for the macroscopic
description of the mixture, in terms of $n_{_i}$, $\bfm u$
and $T$.\\

\sect{Euler approximation} If the characteristic time of
relaxation due to elastic interactions is much shorter than the
one due to the reactions, we can set $Q_{_i}[{\bfm f}]=0,$ whose
solutions are the Maxwell distribution functions
\cite{Cercignani}:
\begin{eqnarray}
f_{_i}({\bfm
v})=n_{_i}\,\left(\frac{m_{_i}}{2\,\pi\,\chi\,T}\right)^{3/2}\,\exp\left[-\frac{m_{_i}}{2\,\chi\,T}\,({\bfm
v}-{\bfm u})^2\right] \ .
\end{eqnarray}
At this point, ${\bfm u}_{_i}$, $I\!\!P$, ${\bfm q}_{_{\rm th}}$
and ${\bfm q}_{_{\rm int}}$, as well as the source terms $S_{_i}$,
can be calculated as follows:
\begin{eqnarray}
{\bfm u}_{_i}={\bfm u} \ ,\hspace{10mm}I\!\!P=p\, I\!\!I \
,\hspace{10mm}{\bfm q}_{_{\rm th}}={\bfm q}_{_{\rm int}}=0 \ ,
\end{eqnarray}
where $p=n\,\chi\,T$ is the pressure and
\begin{eqnarray}
S_{_i}=\lambda_{_i}\Big[\nu_{_{34}}(T)\,n_{_3}\,n_{_4}
-\nu_{_{12}}(T)\,n_{_1}\,n_{_2}\Big]
\ ,\label{g}
\end{eqnarray}
where the effective collision frequencies $\nu_{_{ij}}$ are given
by
\begin{eqnarray}
\nonumber \nu_{_{12}}(T)&=&\left(\frac{\sqrt{m_{_1}\,m_{_2}}}
{2\,\pi\,\chi\,T}\right)^3\,\times\\
&&\int\int\int V_{_{12}}\,I^{^{34}}_{_{12}}(V_{_{12}},\,\zeta)
\,\exp\left(-\frac{m_{_1}\,{\bfm v}^2+m_{_2}\,{\bfm
w}^2}{2\,\chi\,T}\right)\,d{\bfm v}\,d{\bfm w}\,d{\bfm
\Omega}^\prime \
,\label{v12}\\
\nonumber \nu_{_{34}}(T)&=&\left(\frac{\sqrt{m_{_3}\,m_{_4}}}
{2\,\pi\,\chi\,T}\right)^3\,
\times\\
&&\int\int\int
V_{_{34}}\,I^{^{12}}_{_{34}}(V_{_{34}},\,\zeta)\,\exp\left(-\frac{m_{_3}\,{\bfm
v}^2+m_{_4}\,{\bfm w}^2}{2\,\chi\,T}\right)\,d{\bfm v}\,d{\bfm
w}\,d{\bfm \Omega}^\prime \ ,\label{v34}
\end{eqnarray}
and satisfy the following relationship \cite{Rossani}:
\begin{eqnarray}
\frac{\nu_{_{34}}(T)}{\nu_{_{12}}(T)}= {\mathcal
R}\,\exp\left(\frac{\Delta E}{\chi\,T}\right) \ ,\label{arrhenius}
\end{eqnarray}
(the Arrhenius law), where ${\mathcal
R}=(\mu_{12}/\mu_{34})^{3/2}$ and $\Delta E=E_3+E_4-(E_1+E_2)$
(without any loss of generality, we can take $\Delta E>0$). At
chemical equilibrium, $G_{_{12}}=G_{_{34}}$, that is, $S_{_i}=0$.\\
In this work we will consider the following class of microscopic
collision frequencies:
$V_{_{34}}I_{_{34}}^{^{12}}(V_{_{34}},\,\zeta)=C_{_{34}}
(\zeta)\,V_{_{34}}^{\,\,\delta}$ where $C_{_{34}}$ is a certain
function of the $\cos(\zeta)$ and $\delta\leq1$ describes the
interaction law. The cases $\delta=0$ and $\delta=1$ are well
known as Maxwell and hard sphere models, respectively. After some
calculations, from Eq. (\ref{v34}), we explicitly obtain
\begin{eqnarray}
\nonumber \nu_{_{34}}(T)&=&
4\,\sqrt{\pi}\,\left(\frac{2\,\chi\,T}{\mu_{_{34}}}\right)^{\delta/2}
\,\Gamma\left(\frac{\delta+3}{2}\right)\,\int\limits_{-1}\limits^1
C_{_{34}}(\zeta)\,d\zeta\\ &=&A_{_{34}}\,(\chi\,T)^{\delta/2}\
,\label{a}
\end{eqnarray}
where $\Gamma$ is the gamma function \cite{AS}, with
$\delta>-3$.\\ Due to the Arrhenius law, we soon obtain
\begin{eqnarray}
\nu_{_{12}}(T)=\frac{A_{_{34}}}{{\mathcal
R}}\,(\chi\,T)^{\delta/2}\,\exp\left(-\frac{\Delta
E}{\chi\,T}\right) \ ,\label{nu}
\end{eqnarray}
where $A_{_{34}}$ is a constant.
\sect{Dispersion relation}

Let us introduce the following notations to indicate a
homogeneous equilibrium state $n_{_i}\equiv n_{_i}^{\rm e}$,
$T\equiv T^{\rm e}$ with $S_{_i}\equiv S_{_i}^{\rm e}=0$ and we
assume, without any loss of generality, ${\bfm u}=0$.\\ We
consider the propagation of a harmonic plane wave:
\begin{eqnarray}
n_{_i}&=&n_{_i}^{\rm e}+{\mathcal
N}_{_i}\,\exp[i\,(\chi\,x-\omega\,t)]
\ ,\label{p1}\\
u&=&{\mathcal U}\,\exp[i\,(k\,x-\omega\,t)]
\ ,\\
T&=&T^{\rm e}+{\mathcal T}\,\exp[i\,(k\,x-\omega\,t)] \
,\label{p3}
\end{eqnarray}
where $u={\bfm u}\cdot{\bfm i}$, where ${\bfm{i}}$ is the
unit vector along the wave propagation.\\
By inserting Eq.s (\ref{p1})-(\ref{p3}) into Eq.s
(\ref{particle}), (\ref{momentum}) and (\ref{totenergy}), after
the linearization with respect to ${\mathcal N}_{_i}$, ${\mathcal
U}$ and ${\mathcal T}$, we obtain:
\begin{eqnarray}
&&-i\,\omega\,{\mathcal N}_{_1}+i\,\kappa\,n_{_1}^{\rm
e}\,{\mathcal U}=S_{_1}^\ast \ ,\label{ss1}\\
&&-i\,\omega\,\left({\mathcal N}_{_1}-{\mathcal
N}_{_2}\right)+i\,\kappa\,\left(n_{_1}^{\rm
e}-n_{_2}^{\rm e}\right)\,{\mathcal U}=0 \ ,\\
&&-i\,\omega\,\left({\mathcal N}_{_1}+{\mathcal
N}_{_3}\right)+i\,\kappa\,\left(n_{_1}^{\rm
e}+n_{_3}^{\rm e}\right)\,{\mathcal U}=0 \ ,\\
&&-i\,\omega\,\left({\mathcal N}_{_1}+{\mathcal
N}_{_4}\right)+i\,\kappa\,\left(n_{_1}^{\rm
e}+n_{_4}^{\rm e}\right)\,{\mathcal U}=0 \ ,\\
&&-i\,\omega\,\rho^{\rm e}\,{\mathcal U}+i\,k\,\chi\,\left(n^{\rm
e}\,{\mathcal T}+T^{\rm e}\,\sum_i{\mathcal N}_{_i}\right)=0 \ ,\\
\nonumber &&-i\,\omega\,\frac{3}{2}\chi\!\left(n^{\rm
e}\,{\mathcal T}+T^{\rm e}\sum_i{\mathcal
N}_{_i}\right)-i\,\omega\,\sum_iE_{_i}\,{\mathcal N}_{_i}\\&& \ \
+i\,\kappa\left(\frac{5}{2}\,\chi\,n^{\rm e}\,T^{\rm
e}+\sum_iE_{_i}\,n_{_i}^{\rm e}\right)\,{\mathcal U}=0 \
,\label{ss2}
\end{eqnarray}
where
\begin{eqnarray}
\nonumber S_{_1}^\ast&=&\nu_{_{34}}(T^{\rm e})\,(n_{_3}^{\rm
e}\,{\mathcal N}_{_4}+n_{_4}^{\rm e}\,{\mathcal
N}_{_3})-\nu_{_{12}}(T^{\rm e})\,(n_{_1}^{\rm e}\,{\mathcal
N}_{_2}+n_{_2}^{\rm e}\,{\mathcal N}_{_1})\\
&+&{\mathcal T}\,\left[\nu_{_{34}}^\prime(T^{\rm e})\,n_{_3}^{\rm
e}\,n_{_4}^{\rm e}-\nu_{_{12}}^\prime(T^{\rm e})\,n_{_1}^{\rm
e}\,n_{_2}^{\rm e}\right] \ .
\end{eqnarray}
Since at equilibrium $\nu_{_{34}}(T^{\rm e})\,n_{_3}^{\rm
e}\,n_{_4}^{\rm e}=\nu_{_{12}}(T^{\rm e})\,n_{_1}^{\rm
e}\,n_{_2}^{\rm e}=R$, from Eq. (\ref{arrhenius}) it follows that
$\nu_{_{34}}^\prime(T^{\rm e})\,n_{_3}^{\rm e}\,n_{_4}^{\rm
e}-\nu_{_{12}}^\prime(T^{\rm e})\,n_{_1}^{\rm e}\,n_{_2}^{\rm
e}=-R\,\Delta E/\chi\,T^{{\rm e}\,2}$.\\ Non trivial solutions of
the system of Eq.s (\ref{ss1})-({\ref{ss2}) for ${\mathcal
N}_{_i}$, ${\mathcal U}$ and ${\mathcal T}$, are obtained if, and
only if, the determinant of the coefficients vanishes. This
condition gives the dispersion relationship, which links $k$ and
$\omega$ \cite{Rossani}:
\begin{eqnarray}
-\frac{\omega^2}{a_{\rm e
}^2}+k^2-i\,\omega\,\tau\left(-\frac{\omega^2}{a_{\rm f
}^2}+k^2\right)=0 \ ,
\end{eqnarray}
where the coefficients $a_{\rm e}$ and $a_{\rm f}$ are the
equilibrium flow and frozen flow speed of sound, respectively,
while $\tau$ is a characteristic time of the reaction. Explicitly,
we have
\begin{eqnarray}
&&a_{\rm f}^2={5\over3}\,\chi\,{n^{\rm e}\,T^{\rm e}\over\rho^{\rm
e}} \
,\\
&&a_{\rm e}^2=a_{\rm f}^2\left[\sum_i{1\over n_{_i}^{\rm
e}}+{2\over 5\,n^{\rm e}}\left(\frac{\Delta E}{\chi\,T^{\rm
e}}\right)^2\right]\,\left[\sum_i{1\over n_{_i}^{\rm
e}}+{2\over3\,n^{\rm e}}\left(\frac{\Delta E}{\chi\,T^{\rm
e}}\right)^2\right]^{-1} \ ,\\
&&{1\over\tau}=R\left[\sum_i{1\over n_{_i}^{\rm e}}+{2\over
5\,n^{\rm e}}\left(\frac{\Delta E}{\chi\,T^{\rm
e}}\right)^2\right] \ .
\end{eqnarray}
It is easy to verify that $a_{\rm f}>a_{\rm e}$, according to
non-equilibrium thermodynamics \cite{Groot}. We can observe that
the dispersion relation can be written as
$k^2\,P(i\,\omega)+\omega^2\,Q(i\,\omega)=0$ where $P$ and $Q$
are first order polynomials of $i\,\omega$.
\sect{Travelling wave solutions} When the macroscopic fields
$n_{_i},\,u_{_i}$ and $T$ only depend on one spatial coordinate
$x$ and on the time $t$ the macroscopic balance laws in the Euler
approximation become:
\begin{eqnarray}
&&\frac{\partial\,n_{_i}}{\partial\,t}+\frac{\partial}{\partial\,x}(n_{_i}\,u)=S_{_i}
\ ,\label{eq1}\\
&&\frac{\partial}{\partial\,t}(\rho\,u)
+\frac{\partial}{\partial\,x}(\rho\,u^2+n\,\theta)=0 \ ,\\
&&\frac{\partial\,e}{\partial\,t}+
\frac{\partial}{\partial\,x}\,\left[(e+n\,\theta)\,u\right]=0 \ ,
\label{eq3}
\end{eqnarray}
where $\theta=\chi\,T$.\\
In the following we consider the travelling wave solutions of the
system described by Eq.s (\ref{eq1})-(\ref{eq3}), that is, we look
for solutions that only depend on the independent variable
$\xi=x-v\,t$ where, due to the Galilei invariance of the system,
we can set the velocity $v$ of the travelling wave equal to zero.
Eq.s (\ref{eq1})-(\ref{eq3}) become:
\begin{eqnarray}
&&\frac{d}{d\,\xi}\,(n_{_i}\,u)=S_{_i} \ ,\label{eq4}\\
&&\frac{d}{d\,\xi}\,(\rho\,u^2+n\,\theta)=0 \ ,\label{eq5}\\
&&\frac{d}{d\,\xi}[(e+n\,\theta)\,u]=0 \ ,\label{eq6}
\end{eqnarray}
which is a system of six nonlinear ODEs that have to be solved
with the appropriate initial conditions. In the following we use a
plus/minus superscript to indicate the asymptotic values at
$\xi=\pm\infty$. We anticipate that the soliton solution will
connect two asymptotic equilibrium states. This means
\begin{eqnarray}
S_{_i}^{^\pm}=0 \ .\label{equi}
\end{eqnarray}
We fix the values of the density $n_{_i}^-$ and the drift
velocity $u^-$ as initial data, at $\xi\rightarrow-\infty$. Then,
from Eq. (\ref{equi}), using Eq.s (\ref{g})  and
(\ref{arrhenius}), it follows that
\begin{eqnarray}
\theta^{^-}=\Delta E\,\left\{\log\left[{\mathcal R}^{-1}\,
\left(\frac{n_{_1}^-\,n_{_2}^-}{n_{_3}^-\,n_{_4}^-}\right)\right]\right\}^{-1}
\ .\label{temp}
\end{eqnarray}
Taking into account the definition of $S_{_i}$, Eq.s (\ref{eq4})
can be rewritten as follows:
\begin{eqnarray}
&&\frac{d}{d\,\xi}\,(n_{_1}\,u)=S_{_1} \ ,\label{eq7}\\
&&\frac{d}{d\,\xi}\,\left[(n_{_1}-\lambda_{_j}\,n_{_j})\,u\right]=0
\ ,\hspace{20mm}j=2,\,3,\,4 \ .\label{eq8}
\end{eqnarray}
Eq.s (\ref{eq5}), (\ref{eq6}) and (\ref{eq8}) are of immediate
integration and give the constants of motion of the system. Eq.
(\ref{eq7}) is the only {\sl dynamical} equation and allows us to
obtain the structure of the wave after integration.\\
From Eq.s (\ref{eq8}) we obtain:
\begin{eqnarray}
[n_{_1}(\xi)-\lambda_{_j}\,n_{_j}(\xi)]\,u(\xi)
=(n_{_1}^--\lambda_{_j}\,n_{_j}^-)\,u^- \ ,\label{ni}
\end{eqnarray}
and taking into account the definitions of the quantities
$n,\,\rho,\,e_{_{\rm int}}$, given in section 2, we obtain the
following relations:
\begin{eqnarray}
&&n(\xi)=\frac{\mu}{u(\xi)} \ ,\label{n}\\
&&\rho(\xi)=\frac{\varphi}{u(\xi)} \ ,\label{r}\\
&&e_{_{\rm int}}(\xi)=\frac{\epsilon}{u(\xi)}-\Delta
E\,n_{_1}(\xi) \ ,\label{e}
\end{eqnarray}
where $\mu=n^-\,u^-$, $\varphi=\rho^-\,u^-$ and $\epsilon=e_{_{\rm
int}}^-\,u^-+\Delta
E\,u^-\,n_{_1}^-$.\\
By using Eq. (\ref{eq5}) we obtain
\begin{eqnarray}
\theta(\xi)=\frac{u(\xi)}{\mu}\,[\tau-\varphi\,u(\xi)] \
,\label{te}
\end{eqnarray}
with $\tau=\rho^-\,(u^-)^2+n^-\,\theta^-$, which gives the
temperature $\theta(\xi)$ as a function of velocity $u(\xi)$.\\
If we now introduce Eq. (\ref{te}) into Eq. (\ref{eq6}), and use
Eq.s (\ref{n})-(\ref{e}) we obtain, after integration,
\begin{eqnarray}
n_{_i}(\xi)=\frac{\alpha_{_i}}{u(\xi)}+\lambda_{_i}\Big[\beta\,u(\xi)+\gamma\Big]
\ ,\hspace{20mm}i=1,\,\cdots,\,4 \ \label{nn1}
\end{eqnarray}
(it should be observed that $n_i$ is singular for $u=0$), with
$\alpha_{_i}=\lambda_{_i}\,[(\epsilon-\varepsilon)/\Delta
E-(n_{_1}^--\lambda_{_i}\,n_{_i}^-)\,u^-]$,
$\beta=-2\,\varphi/\Delta E$, $\gamma=5\,\tau/2\,\Delta E$ where
$\varepsilon=[\rho^-\,(u^-)^2/2+5\,n^-\,\theta^-/2+e_{_{\rm
int}}^-]\,u^-$. Eq. (\ref{nn1}) gives the concentrations
$n_{_i}(\xi)$ as functions of only the velocity $u(\xi)$.\\
Finally, by using Eq. (\ref{nn1}) in Eq. (\ref{eq7}) follows the
nonlinear ODE:
\begin{eqnarray}
\frac{d\,u}{d\,\xi}=R[u] \ ,\label{difeq}
\end{eqnarray}
with
\begin{eqnarray}
R[u]=\frac{S_{_1}[u]}{2\,\beta\,u+\gamma} \ .\label{R}
\end{eqnarray}
Eq. (\ref{difeq}), after numerical integration, gives us $u(\xi)$
and, if Eq.s (\ref{te})-(\ref{nn1}) are used, we can find
$\theta(\xi)$ and $n_{_i}(\xi)$.\\ Some conditions are now
discussed that have to be satisfied to obtain physically
meaningful results. It is well known that solitons of Eq.
(\ref{difeq}) reach finite asymptotic values which are two
consecutive zeros of $R[u]$ \cite{Drazin}. This implies, as we
anticipated, $S_{_1}^{^\pm}=0$, so that the solitary wave
connects two asymptotic states where the system is at chemical
equilibrium. By choosing the initial data $n_{_i}^->0$ and $u^-$,
we obtain $\theta^{^-}$ from Eq. (\ref{temp}), which must be
positive. This implies the following constraint on $n_{_i}^{-}$
(the first condition):
\begin{eqnarray}
\frac{n_{_1}^-\,n_{_2}^-}{n_{_3}^-\,n_{_4}^-}>{\mathcal R} \ .
\end{eqnarray}
Subsequently, we solve the transcendental equation $S_{_1}[u]=0$.
Its roots are the asymptotic values $u^\pm$ of $u(\xi)$. Regular
solutions of Eq. (\ref{difeq}) require that there are no
singularities of $R[u]$ in the interval $[u_{_{\rm
min}},\,u_{_{\rm max}}]$, where $u_{_{\rm min}}=\min(u^-,u^+)$
and $u_{_{\rm max}}=\max(u^-,u^+)$. Singularities in $R[u]$ arise
both for $u=0$ and $2\,\beta\,u+\gamma=0$. The condition $u\ne0$
implies $u^-\,u^+>0$ (the second condition), while
$2\,\beta\,u+\gamma\not=0$ for $u\in[u_{_{\rm min}},\,u_{_{\rm
max}}]$ is the third condition that has to be satisfied. Finally,
by using Eq.s (\ref{te}) and (\ref{nn1}), from $u^+$ we obtain
the asymptotic values $n_{_i}^+$ and $\theta^{^+}$, which must be
positive (the fourth condition). When all these conditions are
satisfied, by choosing an arbitrary initial value
$u_{_0}\in[u_{_{\rm min}},\,u_{_{\rm max}}]$, we can finally
forward and backward integrate Eq. (\ref{difeq}) in order to
obtain the structure of the travelling wave which connects the
two chemical equilibrium states at $\pm\infty$.
\begin{figure}
\begin{center}
\includegraphics[width=\textwidth]{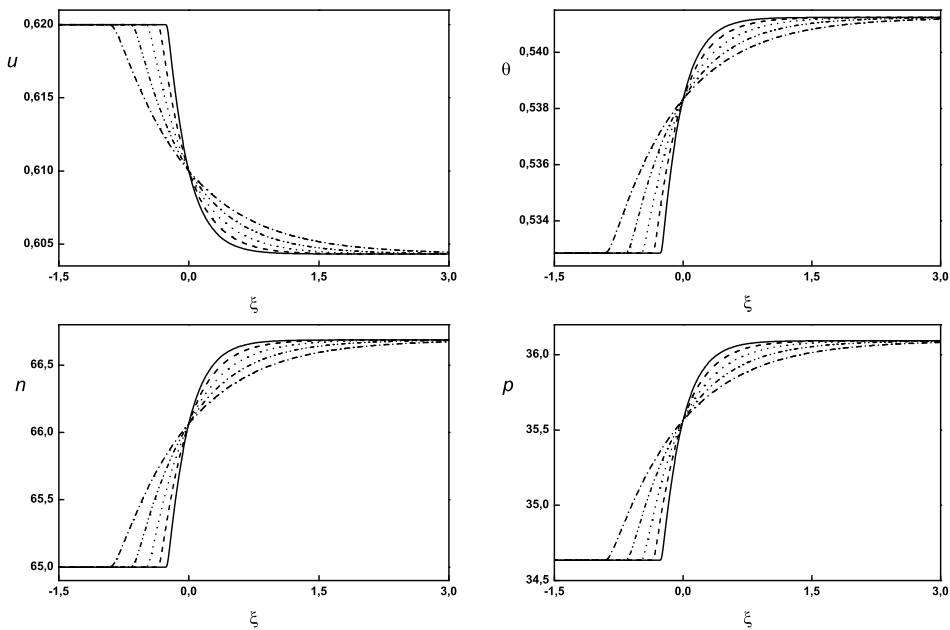}
\end{center}
\caption{Plot, in arbitrary units, of the drift velocity $u$,
temperature $\theta$, total density $n$ and pressure $p$, for
different values of the parameter $\delta$: $\delta=1$ (dash-dot
line), $\delta=0$ (dash-dot-dot line), $\delta=-1$ (dot line),
$\delta=-2$ (dash line) and $\delta\rightarrow-3$ (solid line).}
\end{figure}

In the following we present and discuss some numerical solutions
which connect the same asymptotic equilibrium, for various choices
of $\delta$. The behaviour of the gas is depicted in figure 1 as
a whole, by plotting: the drift velocity $u$, temperature
$\theta$, total density $n$ and pressure $p$. It should be
observed that, as $\theta<1$ in our example, for lower $\delta$
is the higher are $\nu_{_{12}}$ and $\nu_{_{34}}$, and the
steeper the curves. On the other hand, in the case $\theta>1$, we
would get smoother and smoother curves for increasingly lower
values of $\delta$.
\begin{figure}
\begin{center}
\includegraphics[width=\textwidth]{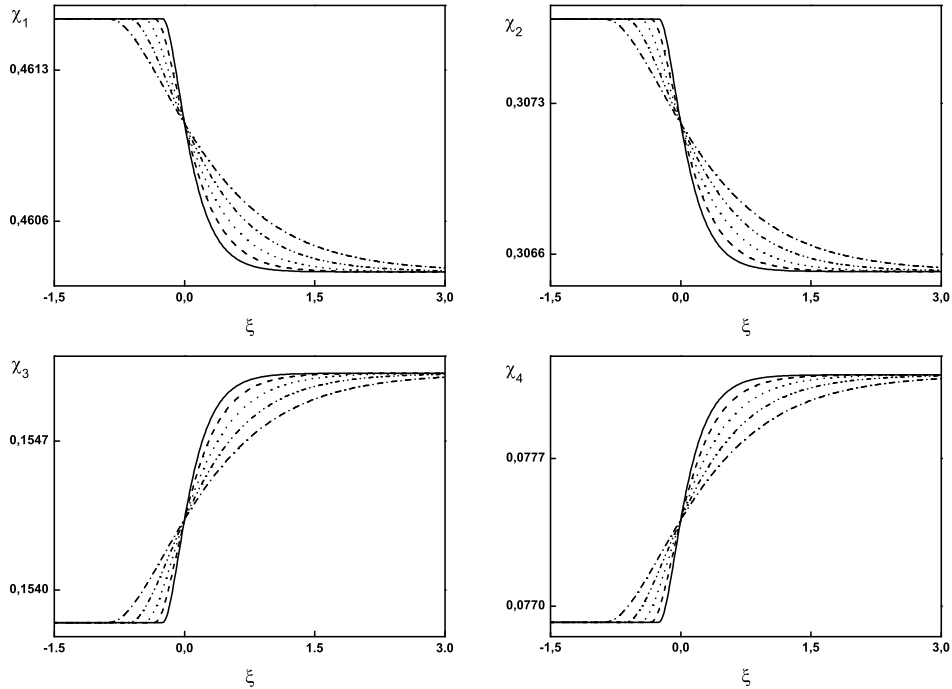}
\end{center}
\caption{Plot, in arbitrary units, of the molar fractions
$\chi_{_i}$, for different values of the parameter $\delta$:
$\delta=1$ (dash-dot line), $\delta=0$ (dash-dot-dot line),
$\delta=-1$ (dot line), $\delta=-2$ (dash line) and
$\delta\rightarrow-3$ (solid line).}
\end{figure}

The molar fractions $\chi_{_i}$ are plotted in figure 2 for the
four species, which are defined by $\chi_{_i}=n_{_i}/n$. This
quantity has another interesting meaning. It is easy to realize
that $\chi_{_i}$ is proportional to the current density
$j_{_i}=n_{_i}\,u$. The slopes
$\chi_{_i}^\prime=d\,\chi_{_i}/d\,\xi$ are simply related by
$\chi_{_1}^\prime=\chi_{_2}^\prime=-\chi_{_3}^\prime=-\chi_{_4}^\prime$,
for each value of $\delta$.\\ Finally, it is possible to observe
that, as the temperature decreases, $\nu_{_{12}}$ decreases with
respects to $\nu_{_{34}}$. This means that the backward reaction
prevails over the forward one, that is, the number of particles 1
and 2 increases while the number of particles 3 and 4 decreases,
as can be seen in the plots.
\sect{Conclusions} We have studied a kinetic model for a mixture
of gases with bimolecular reversible reactions, in the extended
kinetic theory. The kinetic equations have been obtained starting
from the Boltzmann equation and taking into account both the
elastic collision integrals and the chemical collision terms
between the particles. A closed set of equations is obtained in
the Euler approximation, under the assumption that the relaxation
due to the elastic scattering is much quicker than the one due to
the chemical interaction. Six partial differential equations are
derived which govern the evolution of the corresponding unknown
macroscopic fields (the densities $n_{_i}$, momentum and total
energy). Some numerical results have also been presented for the
simple case of a plane steady travelling wave solution, for
various meaningful interaction laws. Finally, the shape of the
soliton that represent the main macroscopic observables for the
gas mixture (the drift velocity, temperature, total density,
pressure and chemical composition) are discussed.

\end{document}